\documentclass[twoside]{article}
\usepackage{fleqn,espcrc2}
\usepackage{graphicx}
\usepackage[figuresright]{rotating}
\newcommand{\AmS}{{\protect\the\textfont2
  A\kern-.1667em\lower.5ex\hbox{M}\kern-.125emS}}
\title{Various regimes of flux motion in Bi$_2$Sr$_2$CaCu$_2$O$_{8+\delta}$ single crystals}
\author{Weimin Chen, J. P. Franck, and J. Jung%
\\Department of Physics, University of Alberta, Edmonton, Canada T6G 2J1}

\begin{document}

\begin{abstract}
Four regimes of vortex motion were identified in the
magnetoresistance of Bi$_2$Sr$_2$CaCu$_2$O$_{8+\delta}$ single
crystals: (1) thermally activated flux flow (TAFF) in samples with
surface defects caused by thermal annealing; (2) TAFF-like plastic
motion of highly entangled vortex liquid at low temperatures, with
$U_{pl} \sim (1-T/T_c)/H^{1/2}$; (3) pure free flux flow above the
region of (2) in clean and optimally doped samples; or, in its
place, (4) a combination of (2) and (3). This analysis gives an
overall picture of flux motion in Bi cuprates.
\end{abstract}

\maketitle

The layered structure of high-$T_c$ cuprates causes intrinsic
pinning even in the vortex liquid state. According to Vinokur {\it
et al.} \cite{vinokur}, such pinning arises from the plastic creep
due to flux entanglement. The highly viscous flux motion gives
rise to an activation-type resistivity, $\rho_{pl} = \rho_0
\exp(-U_{pl}/T)$, with $U_{pl} \sim (1-T/T_c)/H^{1/2}$. This
pinned liquid model had been confirmed in recent experimental
studies\cite{gordeev}. Another interesting phenomenon is free flux
flow, as described by the classical Bardeen-Stephen model
\cite{bardeen}, which occurs when the pinning barrier is
suppressed. Plastic creep and free flux flow are generally
difficult to verify in experiment, because they can be easily
replaced by defect pinning effects. Free flux flow (FFF) was only
observed in clean YBCO samples \cite{Ong}, or under high driving
forces by using current pulses \cite{kunchur}. For Bi cuprates,
intrinsic pinning is weaker due to its higher anisotropy, and is
thus more difficult to investigate. In the present work, we report
different regimes in flux motion under various pinning
circumstances, by measuring the magnetoresistance of
Bi$_2$Sr$_2$CaCu$_2$O$_8$ single crystals.

1. {\it Pinning by surface defects} In annealed samples, thermally
activated flux flow (TAFF) resistance (with $T$-independent
barrier) was consistently observed down to the lowest $T$. Before
annealing, on the other hand, a crossover behavior frequently
occurred: below some crossover $T_x$, plastic creep was identified
in agreement with the pinned liquid scenario; Above $T_x$ it
crossovers to either pure free flux flow or a combination of FFF
and TAFF. This clear difference between as-grown and annealed
samples is connected to the effect of annealing. Scanning
microscopy imaging revealed that the atomic-scale smooth surfaces
of as-grown samples could be damaged by thermal annealing ($\sim
450^\circ$C for 4h). Partial evaporation of material results a
mesh of sub-$\mu$m-sized pits on the sample surfaces. Obviously,
such defects can effectively cause pinning to give rise to the
TAFF behavior over wide temperature ranges, even when the bulk
pinning is absent.

\begin{figure}[t]
\vspace{1pc}
\includegraphics[scale = 0.36]{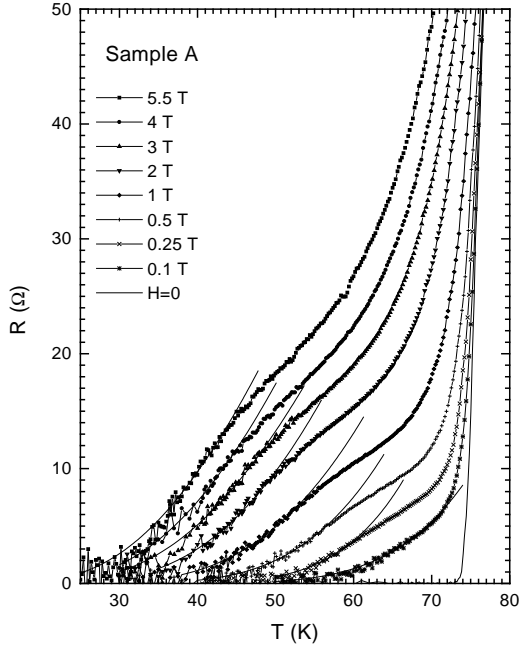}
\vspace{-2pc} \caption{Resistance of an as-grown Bi-2212 sample.
Crossover in flux motion is evident as $T$ increases. The solid
lines are fittings to plastic-flow model: $R_{pl} \propto
exp[-U_{pl}(H,T)/T]$. \label{fig1}} \vspace{-1.2pc}
\end{figure}
\begin{figure}[t] \vspace{1.1pc}
\includegraphics[scale = 0.35]{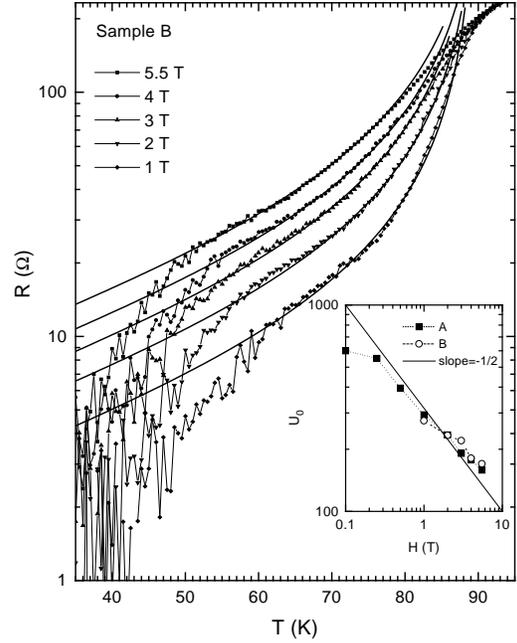}
\vspace{-2pc} \caption{Crossover behavior and free flux flow
fitting (solid lines). Inset: field dependence of $U_0$ (see text)
for Samples A and B, showing an approximate $1/\sqrt{H}$ relation
(straight line). \label{fig2}} \vspace{-1.15pc}
\end{figure}
2. {\it Pinning in vortex liquid} Figure~\ref{fig1} shows a
typical set of resistance data for fields applied along the
$c$-axis of an as-grown Bi-2212 sample ($T_c \sim 76$ K). The
crossover in flux motion is evident. Similar phenomenon is also
clear in other as-grown samples, one of which is presented in
Fig.~\ref{fig2}. We tried to fit the tail below the crossover
temperature to the pinned-liquid model \cite{vinokur}: $\rho_{pl}
= \rho_0 \exp(-U_{pl}/T)$, with $U_{pl} \sim \phi_0^2a(1 -
T/T_c)/8\gamma \pi^2 \lambda_L(0)^2 \equiv U_0(1-T/T_c)$, where $a
\sim (\phi_0/H)^{1/2}$ is the inter-vortex spacing. The result is
given by the solid lines in Fig.~\ref{fig1}, which shows
satisfactory agreement. The extracted field dependence of $U_0$
from Figs.~\ref{fig1} and \ref{fig2} is plotted in the inset of
Fig.~\ref{fig2}. At high fields, $U_0$ also follows the
$1/\sqrt{H}$ dependence as predicted (the straight line of -1/2
slope ). Similar behavior and crossover at low fields were
reported for Bi-2212 \cite{2212} and 2223 \cite{2223}.

3. {\it Free flux flow} Above the crossover temperature, flux
lines become disentangled and the pinning vanishes ($U<k_BT$). In
this regime, free flux flow is thus expected if there exists no
defect pinning. We tried to fit the resistance to the
Bardeen-Stephen model: $R_f=R_NH/H_{c2}$, where $R_N$ is the
normal-state extrapolated resistance and $H_{c2}=\beta(T_c-T)$.
One of the results is as shown in Fig.~\ref{fig2} (solid lines).
The fitting is impressive over a very wide temperature range (51
to 78 K for $H=5.5$ T, $T_c \sim$ 85 K). The value of $\beta$ is
about 0.8 to 1.4 T/K, increasing with field. This result closely
agrees with literature data (0.75 T/K) \cite{dhc2dt}.

4. An intermediate situation involves combined contributions of
both FFF and TAFF, as revealed by resistivity of YBCO
\cite{giura}. The total resistivity is expressed as \cite{review}:
$1/\rho = 1/\rho_{fff} + 1/ \rho_{TAFF}$. A preliminary fitting of
our data to this scheme seems to explain the resistance above the
crossover $T$ in Fig.~\ref{fig1}.

In summary, we emphasize the pinning effect of the surface defects
caused by thermal annealing, which results in activation-type flux
flow even when the bulk pinning is absent. In clean samples,
vortex motion was analyzed in the schemes of pinned liquid
(plastic flow) and free flux flow.

\end{document}